\documentclass[a4paper,fleqn]{cas-dc}

\usepackage[numbers]{natbib}
\usepackage{layouts}

\def\tsc#1{\csdef{#1}{\textsc{\lowercase{#1}}\xspace}}
\tsc{WGM}
\tsc{QE}
\tsc{EP}
\tsc{PMS}
\tsc{BEC}
\tsc{DE}

\begin{document}
\let\WriteBookmarks\relax
\def\floatpagepagefraction{1}
\def\textpagefraction{.001}

\shorttitle{Defect detection in GFRT by X-ray scattering}

\shortauthors{\"O. \"Ozg\"ul et~al.}

\title [mode = title]{Defect detection in glass fabric reinforced thermoplastics by laboratory-based X-ray scattering}


\author[1]{\"Ozgul \"Ozt\"urk}
\cormark[1]
\ead{Oezguel.Oeztuerk@uni-siegen.de}
\affiliation[1]{organization={Physics Department, University of Siegen},
    addressline={Walter-Flex-Str. 3}, 
    city={Siegen},
    postcode={57072}, 
    country={Germany}}
\credit{Data curation, Investigation, Writing - review \& editing}

\author[2]{Rolf Br\"onnimann}
\affiliation[2]{organization={Empa, Swiss Federal Laboratories for Materials Science and Technology},
    city={D\"ubendorf},
    postcode={8600}, 
    country={Switzerland}}
\credit{Resources, Methodology, Writing - review \& editing}
\ead{rolf.broennimann@empa.ch}

\author[1,3]{Peter Modregger}
\affiliation[3]{organization={Center for X-ray and Nano Science, Deutsches Elektronen-Synchrotron},
    city={Hamburg},
    postcode={22607}, 
    country={Germany}}
\credit{Conceptualization, Formal analysis, Supervision, Visualization, Writing - original draft}
\cormark[2]
\ead{peter.modregger@uni-siegen.de}

\cortext[cor1]{Corresponding author}
\cortext[cor2]{Principal corresponding author}

\begin{abstract}
Glass fabric reinforced thermoplastic (GFRT) constitutes a class of composite materials that are especially suited for automobile construction due to their combination of low weight, ease of production and mechanical properties. However, in the manufacturing process, during forming of prefabricated laminates, defects in the glass fabric as well as in the polymer matrix can occur, which may compromise the safety or the lifetime of components. Thus, the detection of defects in GFRTs for production monitoring and a deep understanding of defect formation/evolution is essential for mass production. Here, we experimentally demonstrate that a certain type of defect (i.e., local fiber shifts), can be detected reliably by X-ray scattering based on the edge-illumination principle. This implies applications for research on mechanism of defect formation as well as for industrial application in production monitoring.
\end{abstract}


\begin{highlights}
\item X-ray transmission and scattering signals are anti-correlated for GFRT samples
\item Local fiber shifts defects in GFRT can be detected by X-ray transmission and scattering
\item Defects are detectable in the hidden glass fiber layer
\end{highlights}


\maketitle

\section*{Introduction}

Lightweight composite materials have attracted ever increasing attention from researchers and from the manufacturing industry in recent decades. Their unique combination of properties such as high strength to weight ratio, resistance to corrosion and economical manufacturing render them suitable for applications such as aerospace, automobiles or construction~\cite{Rajak2019}. A class of composite materials are fiber reinforced polymers, which are generally comprised of several fiber layers (e.g., glass, carbon or aramid) nested in a polymer matrix (e.g., thermoplastic, epoxy or vinyl ester)~\cite{Masuelli13}. Each of these material systems has their own unique mechanical properties and area of application~\cite{seydibeyoglu2017}.

Glass fabric reinforced thermoplastics (GFRT) are especially suitable for automobile manufacturing due to its ease of production, high stiffness, corrosion resistance and recyclability~\cite{FAN20163}. Naturally, one the main objectives of substituting metallic components by GFRT is weight reduction, which leads to reduced fuel consumption and ultimately reduced CO$_2$ emissions. Furthermore, GFRT components can be produced by conventional forming presses from ready-made, prefabricated laminates, which is comparatively cost efficient. However, defects in the glass fabric or the thermoplastic matrix can be created or enlarged during forming, which poses a serious safety concern for structural components~\cite{BERRY20179,SAFRI2019133}. Therefore, various non-destructive testing (NDT) techniques are investigated for the understanding of defect formation and evolution as well as for the application in production monitoring of GFRT components~\cite{Gholizadeh2016}. Examples for NDT techniques applied to composite materials include ultrasonic imaging~\cite{Liu2013,Dong2015}, THz imaging~\cite{Cristofani2014,Wang2019,Souliman2022}, X-ray computed tomography~\cite{Wang2019} or infrared thermography~\cite{Yang2016}. 

\begin{figure}
    \centering
    \includegraphics[width=0.45\textwidth]{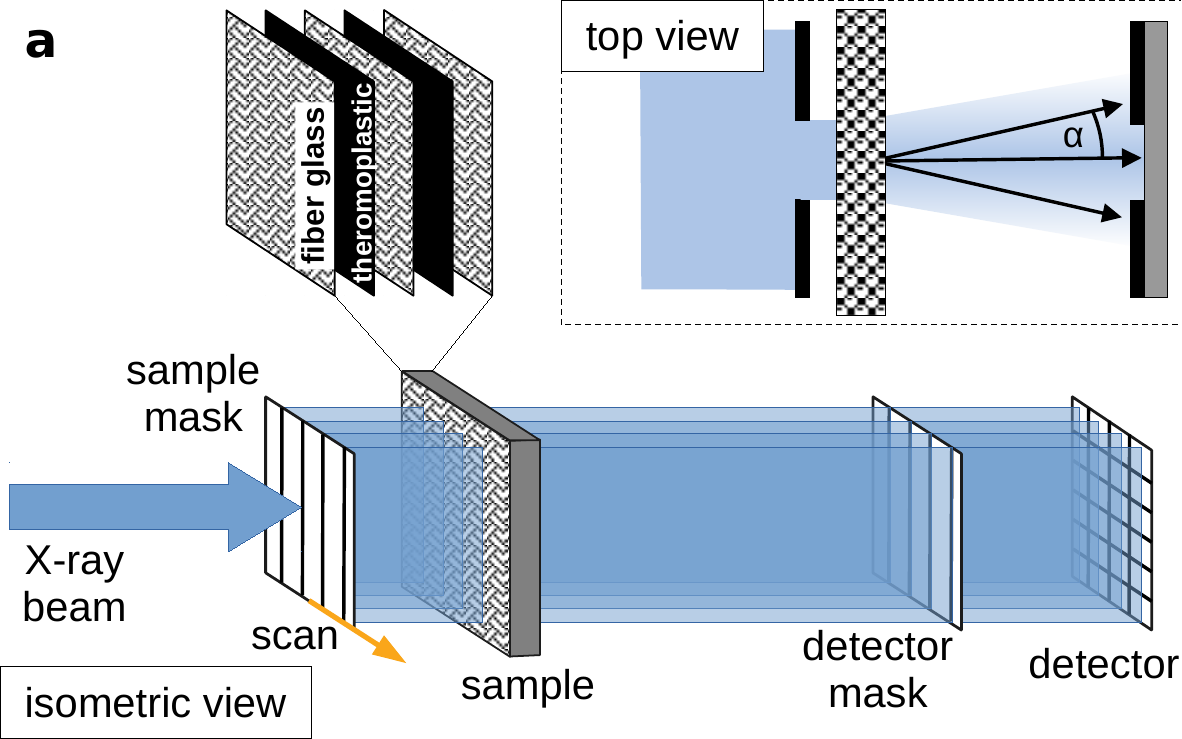}\\
    \vspace{5mm}
    \includegraphics[width=0.45\textwidth]{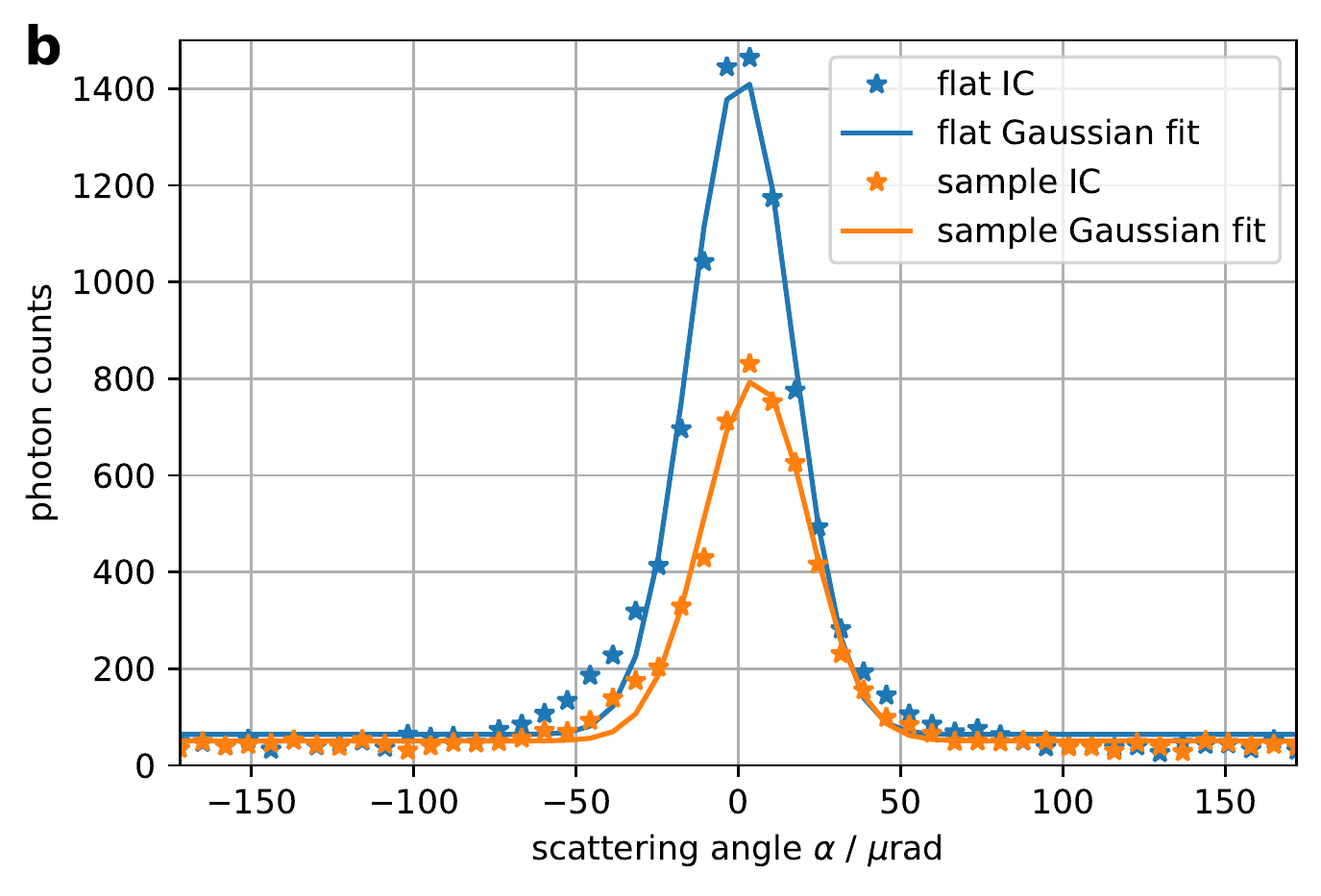}
    \caption{Principle of X-ray imaging with edge-illumination. (a) Sketch of the setup and the composite material system (GFRT). In the experiment, the sample mask is laterally scanned, which provides a Gaussian-like intensity signal, called the illumination curve, for each detector pixel as shown in (b). This panel shows the flat-field IC, measured without sample, and a sample IC, measured with a sample, for the same detector pixel. Intensity reduction is due to absorption, a shift in peak position is due to refraction and a broadening of the curve is due to sub-pixel scattering.}
    \label{fig:fvt_setup}
\end{figure}

In addition, laboratory-based X-ray scattering based on the edge-illumination (EI) principle was applied to defect detection in composite materials~\cite{Endrizzi2015}. EI constitutes a phase-sensitive X-ray imaging technique that simultaneously provides absorption, phase and scattering contrast~\cite{Endrizzi2014d,Endrizzi2014e,Olivo2021}. The principle of image formation can be described as follows (Fig.~\ref{fig:fvt_setup}a). The incident X-ray beam is split into smaller beamlets by an apertured sample mask. The beamlets are distorted by the interaction with the sample downstream. These distortions are then analyzed by an apertured detector mask and measured by a pixelated detector. The pitch of the apertures projected onto the detector matches the pixel size and, thus, one sample mask aperture and one detector mask aperture contribute to the signal in a given detector pixel column. In the experiment, the sample mask is laterally scanned by a fraction of its pitch, which results in a Gaussian-like intensity distribution, called the illumination curve (IC), for each detector pixel (Fig.~\ref{fig:fvt_setup}b). A complete data set contains a reference (i.e., flat-field) IC without a sample present in the beam and a sample IC. The aforementioned sample-induced distortions of the beamlets appear as specific differences between the flat-field and the sample IC: absorption reduces the height, refraction shifts the peak position and scattering is reflected in a broadening of the IC. EI constitutes an incoherent imaging system with typical aperture sizes of $>\!\!10\,\mu$m rendering this setup insensitive to mechanical vi\-bra\-tions and compatible with standard laboratory-based X-ray sources.

In the context of defect detection in composite materials, the scattering contrast is of particular interest. Scattering originates from refraction at numerous microscopic sample inhomogeneities~\cite{Modregger2016a,Endrizzi2017a} (e.g., glass-polymer interfaces), which cannot be resolved individually (Fig.~\ref{fig:fvt_setup} a - top view). Thus, the scattering contrast constitutes a measure of sample inhomogeneity within a detector pixel, which renders it a sub-pixel contrast that is complementary to the absorption and the refraction signal. For example, it has been demonstrated that the scattering contrast (and its higher orders) can be used to measure particle sizes down 8~$\mu$m with 250~$\mu$m pixels~\cite{Modregger2017}. The sub-pixel nature of the scattering contrast opens the possibility for faster scans and/or dose reduction by exploiting larger pixel sizes. X-ray scattering was also combined with tomography revealing sample inhomogeneities in three dimensions~\cite{Endrizzi2017a}. Up to now, multi-modal X-ray imaging with EI was applied to the detection of impact damage in a carbon fibre/epoxy resin laminate in radiographic face-on geometry~\cite{Endrizzi2015} as well as in tomographic edge-on geometry, which revealed delamination and cracks~\cite{Shoukroun2020,Shoukroun2020a}, and to the characterization of porosity in face-on geometry for the same material system~\cite{Shoukroun2021,Shoukroun2021a}.

The so-called dark-field contrast~\cite{Pfeiffer2008c} provided by labora\-tory-based grating interferometry (GI)~\cite{Pfeiffer2006a} is functionally equivalent to the scattering contrast accessible by EI~\cite{Modregger2017}. Compared to EI, GI constitutes a coherent imaging system that relies on the fractional Talbot effect (i.e., coherent Fresnel diffraction at periodic structures). The utilization of (partial) coherence implies the necessity for optical components with small structures (few micrometers), higher requirements for the X-ray source and the lack of contribution of parts of the spectrum to the dark-field signal~\cite{Thuering2013a}. Nevertheless, GI is actively utilized in the field of composite materials and examples include the differentiation of material phases in concrete~\cite{Sarapata2015}, the visualization of delamination in carbon fibre reinforced plastic composites (CFRP)~\cite{Vavrik2015}, the characterization of anisotropic fibre orientation in CFRP and short fiber GFRP~\cite{Plank2015,Prade2017,Morimoto2020}, monitoring of bonded aircraft repairs in CRFP~\cite{Roper2020}, the measurement of fiber length in short fiber composites~\cite{Wang2022}, the characterization of fiber waviness in CFRP~\cite{Glinz2022} as well as first investigations into quantitative interpretation~\cite{Kasai2021}. Recently, it was also demonstrated that local microfiber orientation in full sized polymer-matrix composite components can be characterized by synchrotron radiation-based X-ray scattering tensor tomography~\cite{Kim2022}, which is a variation of GI.

Defects in composite materials include delamination (i.e., separation of fiber layers), consolidation (i.e., variations of the polymer matrix thickness), local fiber shifts (i.e., local agglomeration of fibers in one layer leaving behind a hole in the fabric of this layer), global fiber shift (i.e., lateral offset of one fabric layer with respect to the others) or cracks (i.e., broken fibers in one or more fabric layer). For GFRT it has been demonstrated that delamination and consolidation can be detected by THz imaging~\cite{Souliman2022}. In this study, we are aiming to provide a proof of concept for the detection of local fiber shifts in GFRT by means of X-ray scattering with EI. To this end, we will first describe the experimental setup and then describe the retrieval of the multi-modal contrasts from experimental images. Afterwards, we will show the absorption and scattering contrasts for 9 GFRT samples (3 control, 3 with a local fiber shift in the top layer and 3 with a local fiber shift in the middle layer) and establish an expected negative correlation between these contrasts. We will introduce the standard deviation of transmission and scattering signals over the field of view (FoV) as an observable for automatic defect detection. Finally, we will use a harmonic analysis of vertically averaged scattering signals in order to demonstrate a reliable detection of local fiber shifts in GFRT composites.

\section*{Experiment and contrast retrieval}

The glass fabric of the GFRT samples were made from bi-directional twill 2/2 E-glass roving fiber, where the twill 2/2 indicates that two wrap yarns (horizontal fibers) travel over and under two weft yarns (vertical fibers) forming a diagonal pattern. Three glass fabric layers were incorporated in a thermoplastic polymer matrix (Fig.~\ref{fig:fvt_setup}a) made of polypropylene (PP) and had a thickness of 1.5~mm. The samples were produced by Bond-Laminates, Lanxess Group (Brilon, Germany).

The laboratory-based experiment was carried out with a newly build EI set-up at the University of Siegen (Germany). The source was an XWT-160-SE X-ray tube (X-RAY WorX, Garbsen, Germany) with a tungsten target operated at 0.5~mA current and 40~kV voltage. While it is possible to increase the current, we have observed an unacceptable increase of source size for this tube. The X-ray detector was a Pilatus 3X 200 K-A (DECTRIS, Baden, Switzerland) with a Si absorber and a pixel size of 172~$\mu$m. Both masks were manufactured by laser ablation (EMPA, D\"ubendorf, Switzerland) of tungsten foil with a thickness of 150~$\mu$m (Goodfellow, Huntingdon, UK). The sample mask featured a pitch of 137.6~$\mu$m and line apertures sizes of 17.2~$\mu$m and was placed at 0.41~m away from the detector. The detector mask featured a pitch of 163.4~$\mu$m and line apertures sizes of 23~$\mu$m and was placed at 0.1~m upstream of the detector. The different pitches account for geometric magnification and provided projected pitches at the detector, which matched the pixel size. The sample to detector distance was 0.36~m and total setup length was 2.05~m. The FoV was limited by the size of the test masks to $(2\times1.7)\,\mathrm{cm}^2$. This setup constituted a prototype utilizing the instrumentation at hand and can be significantly optimized (e.g., a more appropriate combination of X-ray source and detector).

In the experiment, the flat-field ICs were measured by scanning the sample mask over one period in 50~steps with an acquisition time of 50~s (Fig.~\ref{fig:fvt_setup}b). The GFRT samples were oriented so that one weave is horizontal and the other is vertical in the detector image. Regions of interest were identified by standard absorption contrast (i.e., masks moved out of the X-ray beam) and then imaged at 50 IC positions using 16 dithering steps with (i.e., moving the sample in sub-pixel steps~\cite{Diemoz2014f}) with a frame time of 3.125~s each. While dithering in EI is generally used to improve spatial resolution, here the corresponding frames were added up to improve the sub-pixel sampling of the composites. Thus, total acquisition time per sample IC step was 50~s and total scan time was about 90~min. 

Retrieval of the multi-modal contrasts provided by EI involved a pixel-wise fit~\cite{Diemoz2010a} of the measured flat-field IC and sample IC to a Gaussian function with an additional offset $c$
\begin{equation}
I_{f,s}(\alpha,x,y) = A_{f,s}(x,y)\,e ^{- \frac{(\alpha-\mu_{f,s}(x,y))^2}{2{\sigma_{f,s}}^2(x,y)}} + c(x,y)
\end{equation}
with $\alpha$, the scattering angle (Fig.~\ref{fig:fvt_setup}a - top view), $(x,y)$, the pixel position, $A_{f,s}(x,y)$, the height of the Gaussian, $mu_{f,s}(x,y)$, the peak position, ${\sigma_{f,s}}^2(x,y)$, the width and the subscripts $f$ and $s$ designate the flat-field and sample ICs, respectively. Exemplary fitting results are shown in Fig.~\ref{fig:fvt_setup}b. The offset $c(x,y)$ had an average value 5\% of the peak over the FoV and was likely due to parasitic transmission through aperture edges or X-ray source tails. Finally, the transmission $t(x,y)$ was calculated according to
\begin{equation}
    t(x,y) = \frac{A_s(x,y)}{A_f(x,y)},
\end{equation}
the refraction $\Delta\alpha$ (no images shown) according to 
\begin{equation}
    \Delta\alpha(x,y) = \mu_s(x,y)- \mu_f(x,y),
\end{equation}
and the sub-pixel scattering signal according to
\begin{equation}
    \sigma^2 (x,y) = {\sigma_s}^2 (x,y) - {\sigma_f}^2 (x,y).
\end{equation}
The resulting contrasts were $2\times 2$ pixel binned to decrease variations over the FoV and resulting in an effective pixel size of 344~$\mu$m.

\section*{Defect detection}

\begin{figure*}[h]
    \centering
    \includegraphics[width=0.99\textwidth]{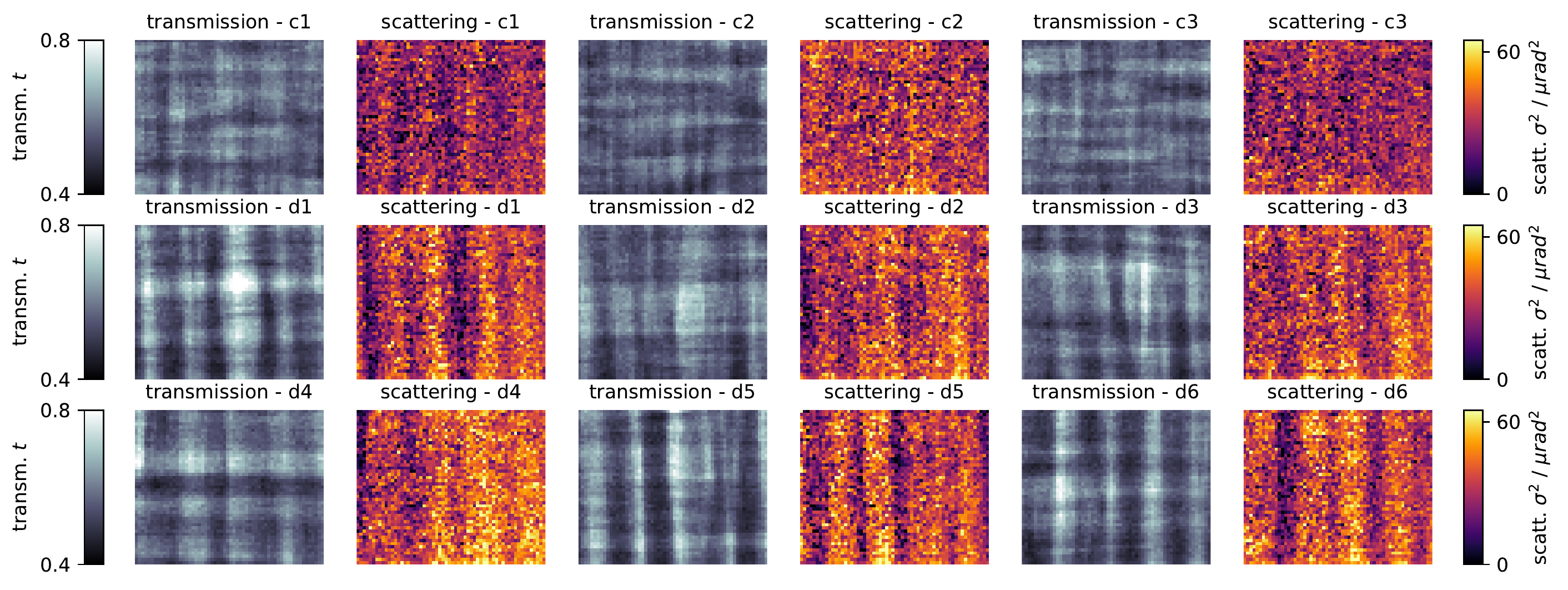}
    \caption{Transmission and scattering contrasts of 9 GFRT samples. All transmission (scattering) images share the same colormap as indicated on the left (on the right) and a FoV of $(2\times1.7)\,\mathrm{cm}^2$ is shown. Top row are the control samples (c1-c3), middle row are the samples with a local fiber shift defect in the top glass fabric layer (d1-d3) and bottom row are the samples with a local fiber shift defect in the middle glass fabric layer (d4-d6).}
    \label{fig:all_constrasts}
\end{figure*}

Figure~\ref{fig:all_constrasts} shows the measured transmission and scattering signals for 3 control samples (c1-c3; top row), 3 samples with a local fiber shift defect in the top fabric layer (d1-d3; middle row) and 3 samples with a local fiber shift defect in the (hidden) middle fabric layer (d4-d6; bottom row). The average transmission value of the control samples ($\bar t_c = 0.570\pm 0.038$; interval is the standard deviation) did not differ significantly from the corresponding value of the defect samples ($\bar t_d = 0.579\pm 0.061$). This indicates that the entire defective area was imaged as the total amount of glass fibers was constant and a partial measurement would lead to larger differences. While the same holds true for the scattering values ($\bar \sigma^2_c = (28\pm 11)\,\mu\mathrm{rad}^2$ versus $\bar \sigma^2_d = (34\pm 13)\,\mu\mathrm{rad}^2$), the latter showed a significantly larger relative variation within the FoV (the 'noisy' appearance of scattering images in Fig.~\ref{fig:all_constrasts}). This is due to the fact that the width of a single glass fiber ($\approx 10\,\mu$m) is comparable to the aperture of the sample mask ($17.2~\mu$m), which results in a strong dependency of the scattering signal on the specific fiber sections exposed to X-rays. In order to lessen the variations in the scattering signal, dithering was used during scanning and pixel binning for data analysis.

Local fiber shifts are visible in the transmission images (Fig.~\ref{fig:all_constrasts}) as distortion of the weave pattern, when visually compared to the control samples. These distortions are also accompanied by an increased variation of transmission values over the FoV. While the latter also holds true for the scattering contrast, local fiber shifts also appear in this contrast as locally increased/decreased scattering values (i.e., increased oscillatory modulations), which outline the defect (especially noticeable for sample d1).

\begin{figure}
    \centering
    \includegraphics[width=0.47\textwidth]{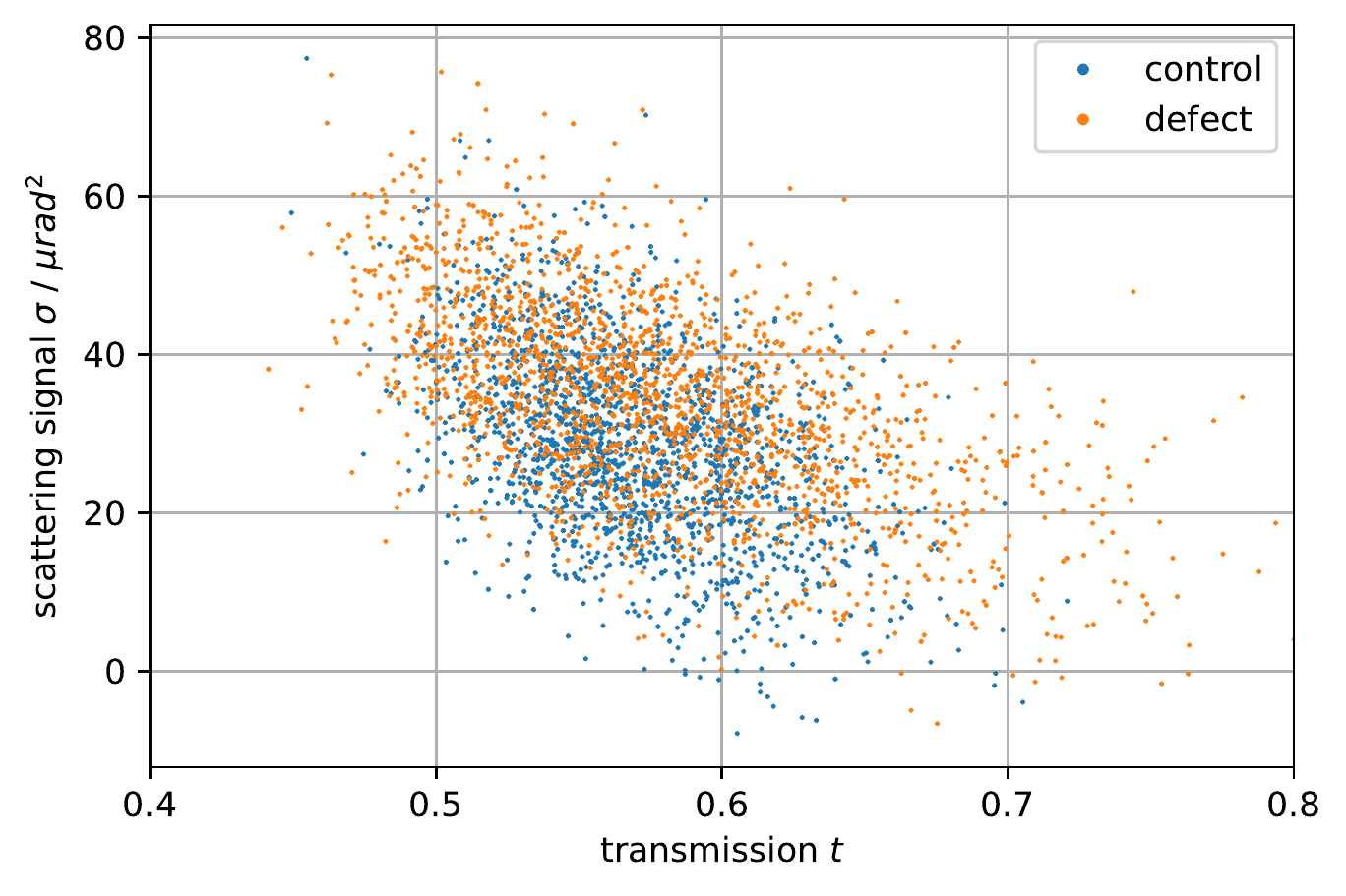}
    \caption{Scatter plot of transmission $t$ and scattering values $\sigma^2$ of 4000 pixels randomly selected from control and defect samples, respectively. The transmission values are negatively correlated (r = -0.52) with the scattering values.}
    \label{fig:abs_sca_corr}
\end{figure}

These observations were consistent with expectations. As stated above, local fiber shifts are local agglomerations of glass fibers in one layer, which leave behind a hole in the fabric of this layer. Since glass (modeled as SiO$_2$ with a density of 2.5~g/cm$^3$) has a much smaller absorption length (1.8~mm~\cite{Chantler2005} at the mean photon energy of 20~keV) than the PP matrix (modeled as C$_3$H$_6$ with a density of 0.86~g/cm$^3$; absorption length is 71~mm~\cite{Chantler2005} at 20~keV), an absence of glass fibers will increase transmission for an identical sample thickness and vice versa for the agglomeration of fibers. Since the scattering contrast originates from glass-PP interfaces, scattering values increase for fiber agglomerations and decrease for absent fibers. These effects are noticeable in Fig.~\ref{fig:all_constrasts}. This explanation is confirmed by the scatter plot of transmission versus scattering values for 4000 randomly selected pixels from all 9 scans shown in Fig.~\ref{fig:abs_sca_corr}. A clear negative correlation (r = -0.52) using all available pixels between the transmission values (e.g., low for fiber agglomerations) and scattering values (e.g., high for fiber agglomerations) demonstrates the agreement. 

\begin{figure}
    \centering
    \includegraphics[width=0.48\textwidth]{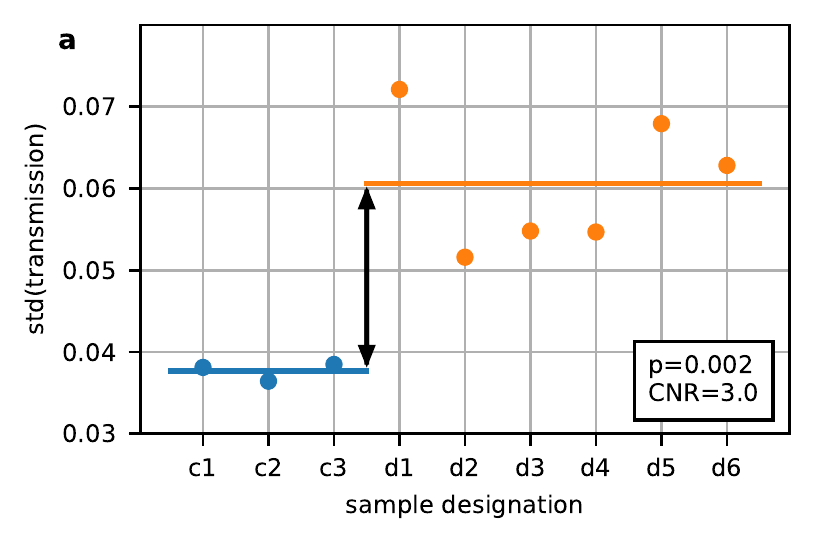}\\
    \includegraphics[width=0.48\textwidth]{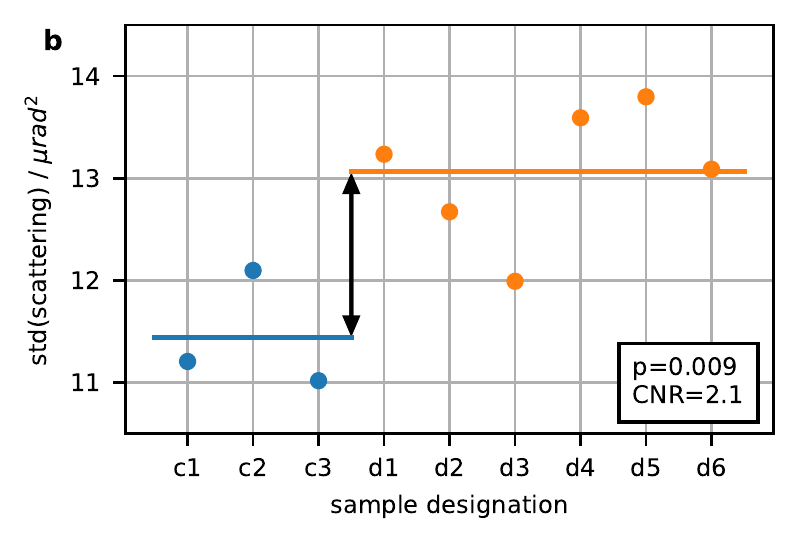}\\
    \caption{The standard deviations of transmission (a) and scattering signals (b) over the FoV in Fig.~\ref{fig:all_constrasts} show a statistically significant difference between control and defect groups for both contrasts. Horizontal lines are group averages.}
    \label{fig:abs_sca_std}
\end{figure}

In the following, we investigated two procedures for automatic defect detection with an application in production monitoring in mind.

First, we exploited the difference in contrast variations between the control and the defective samples. To this end, we used the standard deviation of the 2 contrasts over the FoV as the observable, which is similar to the utilization of the refraction contrast for porosity determination in~\cite{Shoukroun2021a}. Figure~\ref{fig:abs_sca_std} displays the individual standard deviations for the transmission (a) and scattering (b) contrast. We used a two-sided t-test~\cite{montgomery2008design} for the determination of statistical significance of the difference between group means and the contrast to noise ration (CNR), given by
\begin{equation}
    \mathrm{CNR} = \frac{\bar h_1-\bar h_2}{\sqrt{\mathrm{var}(h_1)^2+\mathrm{var}(h_2)}}
\end{equation}
with $\bar h_{1,2}$ the mean of group 1 and 2 and $\mathrm{var}(h_{1,2})$ the corresponding variance, as a measure of effect size.


The results are included in Fig.~\ref{fig:abs_sca_std}, which demonstrate a statistical significant difference between the group means for both contrasts and a robust CNR of 3.0 for the standard deviation of the transmission and of 2.1 for the standard deviation of the scattering contrast. Please note that the defects in the middle layer (samples d4-d6), which are not visually accessible, were as detectable as the defects in the top layer (samples d1-d3). Naturally, the measured standard deviation values depend on the employed FoV. Increasing the FoV for the defective samples to non-defective regions would lower the difference between control and defective observables. However, this can be readily addressed by utilizing a moving image section of appropriate size and highlighting regions of increased standard deviations as potential defects.

Further, taking these results naively at face value one might come to the conclusion that the transmission contrast is superior to the scattering contrast for defect detection. However, using the standard deviation of the transmission contrast as the observable for detection of local fiber shifts implicitly assumes a constant sample thickness as variations in the thickness (i.e., consolidation) could also account for variations in the transmission. In fact, without using the {\em a priori} knowledge that the sample thickness is indeed constant, only the scattering signal may reveal a local agglomeration or absent of glass fibers. Thus, the combination of transmission and scattering contrast has the potential to detect and to distinguish local fiber shifts and consolidation.

\begin{figure}
    \centering
    \includegraphics[width=0.48\textwidth]{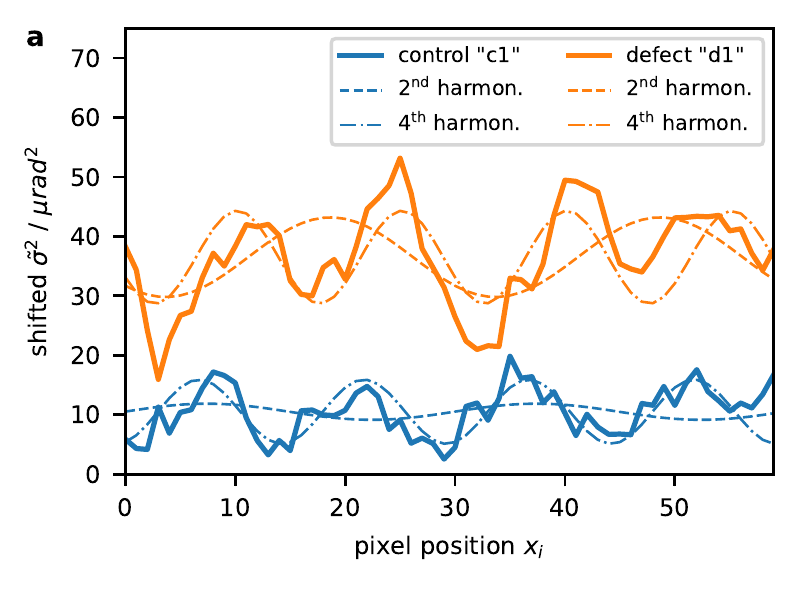}\\
    \includegraphics[width=0.48\textwidth]{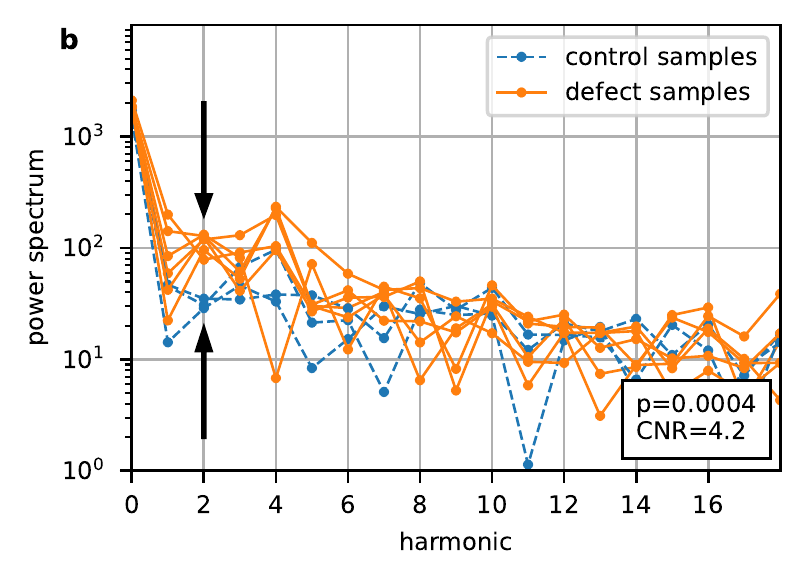}
    \caption{Improved defect detection with scattering. (a) Vertically averaged scattering signals $\sigma^2$ (continuous lines) of control sample 'c1' and defect sample 'd1'. The curves have been shifted for clarity. Dashed lines are cosines corresponding to the 2$^\mathrm{nd}$ and 4$^\mathrm{th}$ harmonics of each vertically averaged scattering signal. The 4$^\mathrm{th}$ harmonic reproduces the periodicity of the glass fiber fabric, while the 2$^\mathrm{nd}$ harmonic (small amplitude for 'c1'; large amplitude for 'd1') is used for defect detection. (b) The power spectra of vertically averaged scattering signals of all samples. A clear distinction in the 2$^\mathrm{nd}$ harmonic between the control and the defect groups is visible.}
    \label{fig:vert_average_harmonic}
\end{figure}

For the reasons laid out above, the second procedure for automatic defect detection investigated used the scattering contrast only. Here, we used a harmonic analysis of the contrast, which suggested itself by the periodic nature of the glass fabric. First, we calculated the vertically averaged scattering values $\tilde \sigma(x_i)$ for the horizontal pixel position $x_i$ by
\begin{equation}
    \tilde \sigma^2(x_i) = \sum_j \sigma^2(y_j,x_i) / N_y
\end{equation}
with $y_i$, the vertical pixel position and $N_y$ the number of pixels in vertical direction. This vertical averaging for increasing the signal is justified by the fact that only horizontal scattering can be observed, which leads to the tendency for vertical features in the scattering contrast (see Fig.~\ref{fig:all_constrasts}). As long as the fiber fabric is roughly oriented along the horizontal/vertical image axes, vertical averaging can be used to increase the signal.

Two examples (1 control and 1 defect sample) for vertically averaged scattering values are displayed in Fig.~\ref{fig:vert_average_harmonic}a, which also illustrates the principle of harmonic analysis~\cite{Press2007}. The cosines associated with the 4$^\mathrm{th}$ harmonic, which reflects the periodicity of the signal, and the cosines associated with the 2$^\mathrm{nd}$ harmonic, which will be used as the observable for defect detection, are shown. As discussed above, a local fiber shift defect increases the scattering signal at the edges of the defect and decreases this contrast in the middle. This is reflected by an increase of the scattering modulation around the defect, which then increases the strength of the 4$^\mathrm{th}$ harmonic. 

Panel~\ref{fig:vert_average_harmonic}b shows the power spectra of the vertically averaged scattering values, which were calculated by numerical Fourier transform. Looking at the 2$^\mathrm{nd}$ harmonic it is evident that the control and the defect group can be readily distinguished. Statistical analysis as above revealed that the CNR is 4.2 (with p=0.0004), which strongly suggests that the scattering signal can be used for a reliable detection of local fiber shift defects in GFRT composite materials. A similar harmonic analysis of the transmission contrasts (not shown) did not lead to any improvement for defect detection.

\section*{Conclusion and outlook}

We have demonstrated that local fiber shifts in GFRT composites can be reliably detected by laboratory-based EI. We have used a face-on sample  geometry for radiographic imaging, which lends well to an application in production monitoring as it is independent from lateral sample sizes. We have introduced the standard deviations over the FoV for the transmission and scattering contrasts as an observable for automatic defect detection and found a robust differentiation between control and defective samples. The CNR between those groups was 3.0 or for the transmission contrast and 2.1 for the scattering contrast. Local fiber shift in the hidden middle layer were as detectable as in the top layer of the GFRT samples. We discussed that the scattering contrast depends only on the number of glass-PP interfaces within a pixel, which renders this contrast significantly less sensitive to total sample thickness (e.g., by consolidation) than the transmission contrast. Further, we have capitalized on the periodicity of the glass fabric and used a harmonic analysis of the vertically averaged scattering signals as an alternative observable. This resulted in an increased CNR between control and defective samples of 4.2. These findings strongly suggest the application of X-ray scattering for research on mechanism of defect formation in composite materials as well as for industrial application in production monitoring.

The study was conducted with a prototype EI setup that can be significantly improved with respect to measurement time and strength of retrieved signals. For example, X-ray tubes of higher power without compromising the source size are available, which in combination with a CdTe-based X-ray detector, would significantly decrease scan times. The scattering signal can be maximized by optimizing aperture sizes of the masks for the application at hand. Finally, measurement times can be further decrease by utilizing a single shot approach, which images the sample on the highly sensitive slope of the ICs and eliminates the necessity for scanning.

\section*{Declaration of competing interest}
The authors have no competing interest to declare.

\section*{Acknowledgements}
\"O. \"Ozt\"urk acknowledges funding by DFG GU $535/6-1$. We would like to thank Matthias Kahl, Aya Souliman and Peter Haring Bol\'{i}var (Institute for High Frequency and Quantum Electronics, University of Siegen, Germany), Bernd Engel (Institute for Production Engineering, University of Siegen, Germany) and Bond-Laminates, a company of the LANXESS Group, for providing samples and reference measurements, and for fruitful discussions.

\printcredits

\bibliographystyle{model1-num-names}

\bibliography{references}


\end{document}